\documentclass[a4paper,fleqn]{cas-dc}

\usepackage[numbers]{natbib}
\begin{document}
\let\WriteBookmarks\relax
\def\floatpagepagefraction{1}
\def\textpagefraction{.001}
\newcommand{\rev}[1]{#1}

\shorttitle{SBHs and NSCs with SHARP}
\shortauthors{Dalla Bont\`a et al.}

\title[mode=title]{Supermassive black holes and nuclear star clusters in galaxies: \\ constraining formation and co-evolution with SHARP and MICADO}

\author[1,2,3]{E. Dalla~Bont\`a}
\cormark[1]
\fnmark[1]\ead{elena.dallabonta@unipd.it} 
\author[4]{E. Portaluri}
\author[2]{C. Arcidiacono}
\author[2]{E. Bortolas}
\author[1,2]{S. Ciroi}
\author[1,2]{E.~M. Corsini}
\author[5]{D. Gasparri}
\author[5]{L. Morelli}
\author[1,2]{A. Pizzella}
\author[6]{M. Signorini}

\affiliation[1]{organization={Dipartimento di Fisica e Astronomia ``G. Galilei”, Universit\`a di Padova, Vicolo dell’Osservatorio 3}, postcode={I-35122},
city={Padova},
country={Italy}}
\cortext[cor1]{Corresponding author}
\affiliation[2]{organization={INAF -- Osservatorio Astronomico di Padova, Vicolo dell’Osservatorio 5}, postcode={I-35122},
city={Padova},
country={Italy}}
\affiliation[3]{organization={Jeremiah Horrocks Institute, University of Central Lancashire}, postcode={PR1 2HE},
city={Preston},
country={UK}}
\affiliation[4]{organization={INAF -- Osservatorio Astronomico d’Abruzzo, Via M. Maggini snc}, postcode={I-64100}, city={Teramo}, country={Italy}}
\affiliation[5]{organization={Instituto de Astronom\'ia y Ciencias Planetarias, Universidad de Atacama, Copayapu 48}, postcode={I-64100}, city={Copiap\'o}, country={Chile}}
\affiliation[6]{organization={ ESA European Space Research and Technology Centre (ESTEC)},
            addressline={Keplerlaan 1}, 
            city={Noordwijk},
            postcode={2201 AZ}, 
            country={the Netherlands}}

\begin{abstract}
Nuclear star clusters and supermassive black holes are the dominant compact components in galaxy nuclei. They can coexist within the same system, and their relative prominence traces the history of gas inflow, star formation, and dynamical evolution on parsec scales. Key questions remain open concerning the physical channels that assemble nuclear star clusters, the mass regimes in which supermassive black holes become dynamically dominant, and the conditions under which compact stellar nuclei are suppressed, destroyed, or replaced by cusp-like structures.

We describe how near-infrared integral-field spectroscopy with SHARP on the Extremely Large Telescope, combined with high-resolution imaging from MICADO, enables quantitative tests of nuclear formation and evolution scenarios within a unified framework. The joint analysis of stellar dynamics, stellar populations, and nuclear structure allows black-hole masses, internal kinematics of nuclear star clusters, and signatures of gas inflow to be measured consistently within the same systems, and defines the mass and distance ranges where robust constraints can be obtained.

Recent discoveries of compact active nuclei at high redshift indicate that efficient black-hole growth can occur in systems with weak or compact stellar components. Although nearby galaxies are not evolutionary analogues of early systems, the diversity of nuclear configurations accessible with SHARP and MICADO provides a local laboratory for isolating the physical mechanisms that regulate the coupled evolution of central black holes and nuclear stellar structures.
\end{abstract}

\begin{highlights}

\item SHARP probes how supermassive black holes and nuclear star clusters assemble.
\item We define diagnostics of in situ formation, cluster infall, and black-hole evolution.
\item SHARP and MICADO measure dynamics, populations, and structure in nearby nuclei.
\item We identify mass and distance regimes where nuclear dynamics can be constrained.

\end{highlights}


\maketitle

\section{Introduction}
The central few tens of parsecs of galaxies host extremely dense stellar systems and, in many cases, supermassive black holes (SBHs, \rev{$M_\bullet \gtrsim 10^6\,M_\odot$, with $M_\bullet$ the black hole mass}). These nuclear components encode the record of gas inflows, star formation, and dynamical evolution that galaxies have experienced \rev{\citep[e.g.][]{Neumayer2020,KormendyHo2013}}.

Early work suggested a simple transition between nuclear star clusters (NSCs) in \rev{low-mass galaxies (with masses $\lesssim10^{10}\,M_\odot$)} and SBHs in \rev{massive galaxies (with masses $\gtrsim10^{11}\,M_\odot$)}, with both classes of objects following similar correlations with host galaxy mass \citep[e.g.][]{Ferrarese2006}. More recent observational studies and theoretical work have revealed a more complex picture. NSCs and SBHs can coexist in the same galaxy, their relative importance depends on galaxy mass and morphology, and multiple formation channels likely operate \citep[e.g.][]{Neumayer2020,KormendyHo2013}.

At the same time, the most massive galaxies often show central light deficits relative to outer S\'ersic profiles, the so called cores \citep[]{Faber1997,Lauer2005}. These are generally interpreted as the outcome of dynamical interaction between supermassive black hole binaries and stars during galaxy mergers, which eject stars on radial orbits and carve out a core \citep{MiloMerritt2001,Merritt2006,Bortolas2021}. 
Stellar dynamical studies with adaptive optics assisted integral field units have provided direct evidence that the orbital structure in core galaxies is consistent with this scenario \citep{Thomas2014}. \rev{Alternative interpretations involving merger-driven violent relaxation and degeneracies in dynamical modelling have also been discussed \citep{Gebhardt2011,KormendyHo2013}, and not all core galaxies show the same degree of tangential anisotropy \citep{Mehrgan2019}.}
 \rev{SBHs with masses $M_{\rm BH}\gtrsim10^{9}\,M_\odot$} have been detected in some of these systems \citep[e.g.][]{DallaBonta2009,Thomas2016,deNicola2025}. However, it remains unclear whether NSCs can form or survive in such environments, and whether nuclear cusp like structures can be rebuilt after core formation.

Recent high resolution simulations have explored the coupled evolution of NSCs and SBHs in low mass galaxies. These models show that the presence of a NSC enhances gas inflow to the central few parsecs, boosting both nuclear star formation and the accretion-driven growth of the black hole \citep{Partmann2025}. Other work has shown that gravitational-wave recoil can induce the formation of nuclear cusps in massive core ellipticals: after the merger, the recoiling black hole can retain a bound cluster of stars and drag them back through the core during its damped oscillations, producing a compact nuclear overdensity \citep{Khonji2025}.

\rev{To fully exploit this information, observations capable of probing the kinematics and stellar structure of galaxy nuclei on parsec scales are required. The combination of SHARP and MICADO on the ELT is ideally suited to this task, providing complementary spectroscopic and imaging observations of nearby galaxy nuclei. Together, these facilities will enable quantitative tests of nuclear formation and evolution scenarios in regimes that are currently inaccessible.}

\rev{The remainder of the paper is organised as follows.} Section~2 summarises the main proposed formation channels, while \rev{Section~3 discusses how the physical structure of galaxy nuclei changes across galaxy mass and motivates a unified observational approach. Section~4 reviews} the current observational constraints, identifies the key observables that can discriminate between scenarios, and describes how SHARP and MICADO can obtain them. 
\rev{Finally, Section 5 discusses the scientific return and observational feasibility of the proposed programme.}

\section{Formation channels for nuclear components}
\subsection{In situ star formation}
One important formation channel for NSCs is in situ star formation from gas that flows into the central few parsecs. In gas rich late type galaxies, bars, spiral arms, and disk instabilities can efficiently drive gas towards the nucleus \citep[e.g.][]{KormendyKennicutt2004}, where it can cool and form stars in repeated episodes. Observationally, many NSCs in late type galaxies show evidence for multiple stellar populations with a wide range of ages, including young components below a few hundred million years \citep[e.g.][]{Walcher2005,Walcher2006,Neumayer2020}. This naturally arises if gas accretion and star formation proceed in bursts.

In situ formation is also possible in early type galaxies when gas is supplied by minor mergers or by stellar mass loss \citep[e.g.][]{Sarzi2006}. The gas originating from stellar mass loss is generally kinematically aligned with the stars and morphologically regular, whereas externally accreted gas is often misaligned or disturbed \citep{Sarzi2006}. In situ formation from enriched gas can produce NSCs that are more metal rich than the surrounding stellar population.
In late type galaxies, in situ formed nuclei can be disky or flattened, especially when angular momentum is conserved during inflow \citep[e.g.][]{Seth2006,Seth2008}.

\subsection{Infall of star clusters}
A second formation channel involves the orbital decay and merging of globular clusters or other massive clusters that spiral into the galaxy centre due to dynamical friction \citep[e.g.][]{CapuzzoDolcetta1993,Antonini2013}. This process is expected to be efficient in galaxies with relatively low velocity dispersions and extended cluster systems. The infalling clusters deposit their stars in the nucleus, building up a dense stellar system over time. This scenario tends to produce NSCs that are dominated by old stellar populations, with metallicities that reflect those of the progenitor clusters, which may be lower than the mean metallicity of the galaxy at the same radius \citep[e.g.][]{Neumayer2020}.

The presence of a SBH strongly suppresses the cluster–infall mechanism: clusters are tidally disrupted before reaching the nucleus when the central black hole exceeds $\sim 10^6\text{--}10^7\,M_{\odot}$, so this channel is effective only in low–mass galaxies \citep{Antonini2013}.

In systems where both gas inflow and cluster infall operate, the resulting NSC may show a mixed origin, with an old, possibly more metal poor component from infall and a younger, more metal rich component from in situ star formation \citep{Neumayer2020}. The relative importance of the two channels likely depends on galaxy mass, morphology, and environment.

\subsection{Black hole dragging and nuclear cusps}

As discussed in the Introduction, massive early-type galaxies with cores are shaped by the dynamical evolution of SBH binaries. Gravitational-wave recoil following SBH coalescence provides an additional pathway for generating compact nuclear structures in these systems \citep{Khonji2025}.

Such cusps differ from classical NSCs in their formation mechanism and are expected to show stellar-population properties broadly consistent with the pre-existing central stars, rather than signatures of recurrent or in situ star formation. They therefore represent a black-hole-driven pathway for generating compact nuclear structure in galaxy nuclei, particularly in massive systems where classical NSC formation channels are inefficient or suppressed.

\section{Physical structure of galaxy nuclei}

The physical connection between NSCs and SBHs depends strongly on host galaxy mass. In low-mass galaxies, high-resolution simulations show that a dense NSC can profoundly influence nuclear gas retention, star formation, and black-hole growth \citep{Partmann2025}. In these models, most of the nuclear cluster mass forms in situ within the central few tens of parsecs, while the black hole grows through intermittent episodes of gas accretion, as young stars and supernovae in the nuclear cluster inject energy and temporarily disrupt the inflow. Without a nuclear cluster, the same black hole seed would grow much less efficiently.

At higher galaxy masses, the conditions that enable such gas-regulated co-evolution become progressively less favourable. In massive galaxies, the SBH is large enough to tidally disrupt infalling clusters before they reach the nucleus, \rev{rendering the cluster-infall channel ineffective}. Moreover, repeated mergers excavate stellar cores through the dynamical interaction of black-hole binaries, inhibiting the formation or long-term survival of classical NSCs \citep[e.g.][]{Antonini2013,Neumayer2020}. As a result, the SBH grows largely independently of any compact stellar nucleus, and nuclear structure is shaped primarily by merger-driven black-hole dynamics rather than by mutual regulation between stars and gas.

Across galaxy mass, the interplay between nuclear components therefore ranges from true co-evolution in low-mass systems to black-hole-dominated nuclear shaping in massive early types. In the context of our SHARP science case, this motivates a unified observational approach that spans the full mass range, enabling us to connect the physical mechanisms that operate \rev{across different galaxy masses and nuclear environments}.

\begin{figure*}
    \centering
    \includegraphics[width=\textwidth]{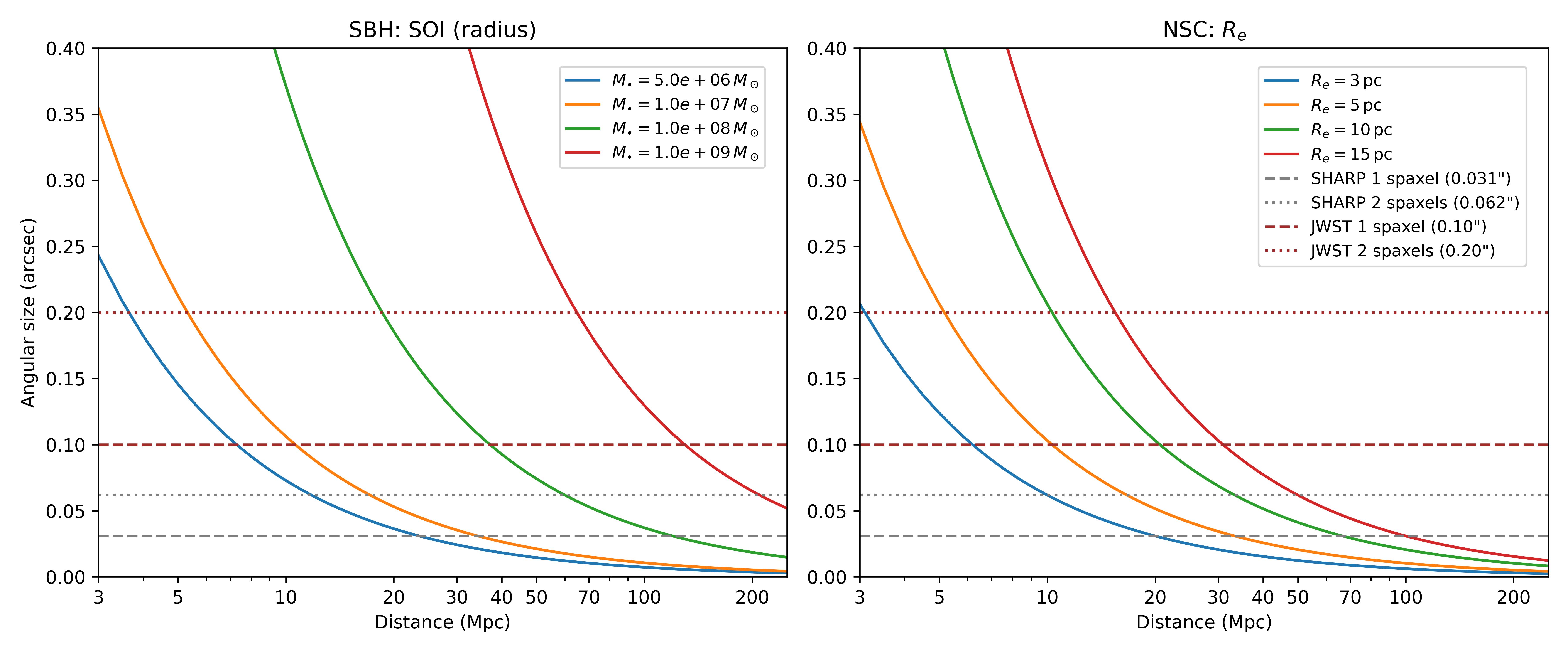}
    \caption{Angular size corresponding to the SBH sphere-of-influence (SOI) radius (left panel) and to NSCs effective radii (right panel) as a function of distance. Solid curves in the left panel correspond to SBH masses of $5\times10^6$ (blue), $10^7$ (orange), $10^8$ (green), and $10^9$ (red) solar masses. \rev{The corresponding SOI radii are $\sim4$, 5, 18, and 60 pc for SBH masses of $5\times10^6\,M_\odot$, $10^7\,M_\odot$, $10^8\,M_\odot$, and $10^9\,M_\odot$, respectively.} Solid curves in the right panel correspond to NSCs effective radii of 3 (blue), 5 (orange), 10 (green), and 15 (red) parsecs. Horizontal lines show the angular sampling of SHARP (1 and 2 spaxels) and of JWST NIRSpec (1 and 2 spaxels), illustrating the mass and distance ranges where SHARP resolves nuclear components.}

    \label{fig:euclid_double_panel}
\end{figure*}

\section{Understanding the co-evolution of SBHs and NSCs with SHARP-VESPER}

The physical mechanisms that regulate the growth of SBHs operate on spatial scales of only a few parsecs. Gas inflow, recurrent nuclear star formation, dynamical heating associated with SBH binaries, the assembly or destruction of NSCs, and the possible formation of recoil-induced cusps all shape the environment in which an SBH can accrete and evolve. 
Disentangling these processes requires spatially resolved spectroscopy at angular resolutions far beyond those achievable with current facilities. \rev{The relevant parsec-scale angular regimes accessible with SHARP-VESPER, operating in the 1.2–2.4~$\mu$m window, are illustrated in Fig.~\ref{fig:euclid_double_panel} and discussed quantitatively in Section~4.1.}

\rev{MICADO will provide diffraction-limited near-infrared imaging with an angular resolution of $4\,$--$12$ mas from the J to K bands \citep{Davies2021}}, suitable for detailed photometric decompositions of galaxy surface-brightness distribution \citep[e.g.][]{DallaBonta2018,Portaluri2014}, while SHARP will deliver near infrared integral field spectroscopy with multiplexing capabilities.

\rev{The synergy between SHARP and MICADO provides a unique view of galaxy nuclei.
SHARP spectroscopy provides the kinematic measurements required to determine SBH masses, NSCs dynamic properties and stellar-population analysis, while MICADO imaging enables the photometric decomposition of the galaxy surface-brightness distribution required to constrain the stellar gravitational potential and study the structural properties of NSCs}.

With these capabilities, SHARP-VESPER enables a physically unified view of SBH–NSC co-evolution. It reveals whether nuclear gas inflow is ongoing or suppressed, whether recent star formation has contributed to the mass budget of the NSC and potentially fuelled the SBH, and whether the dynamical configuration of the nucleus favours or inhibits further accretion. At the same time, the spatially resolved stellar kinematics constrain the gravitational contribution of both the NSC and the SBH, clarifying how their relative dominance varies with host galaxy mass.

By coupling these diagnostics within a single observational framework, SHARP-VESPER establishes the empirical basis for connecting the present-day diversity of nuclear structures to the physical processes that regulate SBH growth and NSC assembly. This connection is essential for identifying the mass range in which co-evolution between SBHs and NSCs is efficient, the regime in which the SBH becomes the dominant component in the nuclear potential, and the nuclear configurations that may share physical characteristics with those expected to enable rapid SBH growth in the early Universe. 

\subsection{Dynamical mapping of SBHs and NSCs}
\label{mapping}
A key requirement for quantifying SBH–NSC co-evolution is the ability to measure the dynamical masses of both components within the same nucleus. The near-infrared spectral range accessible to SHARP-VESPER contains several strong absorption features, including the CO bandheads (2.29–2.40 $\mu$m), the Na{\sc i} doublet at 2.21 $\mu$m, the Ca{\sc i} triplet at 2.26 $\mu$m, and a number of metal lines in the H band (1.50–1.80 $\mu$m). These features are sensitive to stellar temperature, metallicity, and surface gravity in unresolved populations \citep[e.g.][]{Cesetti2009,Lyubenova2010,Riffel2015}, and enable the recovery of velocity, velocity dispersion, and higher-order \rev{line-of-sight velocity distribution} (LOSVD) moments. These measurements form the foundation of orbit-based dynamical modelling \citep[e.g.][]{Cappellari2016}, which requires spatially resolved sampling of the nuclear potential.

For SBHs, the limiting factor is the angular size of the radius of influence, \rev{defined as the region where the gravitational potential of the SBH dominates over that of the surrounding stars \citep[e.g.][]{FerrareseFord2005}}. Reliable dynamical measurements require that this radius be sampled by at least two spaxels. As shown in Fig.~\ref{fig:euclid_double_panel}, this restricts robust SBH mass determinations to galaxies hosting $10^{7}\,M_{\odot}$ black holes within \rev{$\sim$15--20 Mpc}, while systems with SBH masses of $10^{9}\,M_{\odot}$ remain accessible out to \rev{at least $\sim200$ Mpc}.

Adaptive-optics assisted IFU studies have demonstrated that full LOSVD recovery in the near-infrared is essential to avoid biases in SBH mass determinations \citep[e.g.][]{Thomas2014,Nguyen2018}. \rev{As illustrated in Fig.~\ref{fig:euclid_double_panel}, the main advantage of SHARP over current facilities lies not only in reaching parsec-scale resolution, but also in providing the spatial sampling required for robust dynamical measurements of SBHs and NSCs across a broader range of masses and distances.}
Clusters with effective radii of 3–5 pc can be dynamically resolved only within $\sim$10–20 Mpc, whereas larger NSCs with $r_{\rm eff}\gtrsim 10$ pc remain resolved out to 40–50 Mpc. 
These sizes are typical of NSCs in \rev{intermediate-mass galaxies (with masses $\sim10^{10}$--$10^{11}\,M_\odot$)} \citep[e.g.][]{Boker2004,Walcher2005,Neumayer2020}, and they define the range in which SHARP-VESPER can recover internal kinematics, including rotation curves and central velocity dispersions \citep[e.g.][]{Seth2010}.

Photometric identification extends well beyond this limit. Many NSCs exhibit high central surface brightnesses and compact light profiles that remain distinguishable from the surrounding bulge even when their effective radii fall below the SHARP spaxel scale.
MICADO imaging provides the depth and angular resolution needed to identify such compact nuclei and separate them from the underlying stellar component \rev{\citep[]{Arcidiacono2020,Arcidiacono2022}}. 

Recent studies have revealed that nuclei in nearby galaxies frequently host multiple stellar components or nuclear discs \citep[e.g.][]{Morelli2020,Morelli2025,Gasparri2021,Gasparri2024}, highlighting the importance of high-resolution photometry for distinguishing NSCs from embedded discs or pseudo-bulges. This capability is essential for establishing the incidence of NSCs across different host-galaxy and SBH masses, including regimes where the SBH dominates the central potential or where the NSC is intrinsically small.
In systems where both photometric identification and dynamical resolution are achievable, SHARP-VESPER delivers two-dimensional kinematic maps of the NSC. These data provide dynamical NSC masses, rotational properties, and central dispersions, and enable the relative contributions of the SBH, the NSC, and the surrounding bulge to be disentangled with minimal degeneracy.

The near-infrared spectral coverage of SHARP-VESPER provides the diagnostics needed to distinguish among the formation pathways outlined in Section~2, without relying solely on integrated nuclear spectra. Age- and metallicity-sensitive absorption features in the H and K bands—including Na\,{\sc i} at 2.21 $\mu$m, Ca\,{\sc i} at 2.26 $\mu$m, the CO bandheads, and metal lines in the 1.50–1.80 $\mu$m range—allow SHARP to constrain stellar-population signatures associated with extended or recurrent star formation expected in nuclei assembled in situ. 

The detection of nebular and molecular gas tracers within the same wavelength range (Pa$\alpha$ 1.875 $\mu$m, Br$\gamma$ 2.166 $\mu$m, and H$_2$ 1–0 S(1) at 2.12 $\mu$m) further constrains ongoing or recent inflow-driven star formation \citep[e.g.][]{Davies2007,MuellerSanchez2006,Riffel2006}.
In contrast, nuclei assembled predominantly through cluster infall are characterised by uniformly old stellar populations, lower metallicities, and an absence of ionised or molecular gas. These properties can be isolated through the same NIR absorption features, combined with the pressure-supported stellar kinematics obtained from SHARP’s full LOSVD measurements \citep[e.g.][]{Seth2010,Nguyen2018}.

Recoil-induced stellar cusps \rev{\citep{Khonji2025}}, produced by the oscillatory motion of a recoiling SBH remnant, are expected to be nearly spherical and to show limited rotation and weak stellar-population gradients, reflecting their assembly from pre-existing stars rather than from fresh gas. Observationally, they may be associated with a centrally concentrated increase in velocity dispersion, although this signature is not unique and must be interpreted in combination with the nuclear light profile and the full two-dimensional stellar kinematics.
The spectral resolution and spatial sampling of SHARP-VESPER enable this signature to be detected even at distances where previous facilities could not resolve the underlying nuclear structure.

\rev{Even when the sphere of influence or the characteristic nuclear scale is sampled by fewer than two SHARP spaxels (Fig.~\ref{fig:euclid_double_panel}), the combination of SHARP-VESPER central kinematics and high-resolution MICADO photometry still enables the identification of SBH-dominated nuclei and a census of compact nuclear structures in nearby galaxies.}

\subsection{Observational signatures of nuclear regimes across galaxy mass}

The physical connection between SBHs and NSCs is expected to vary across galaxy mass, but observational constraints on the different nuclear regimes remain limited. The relative roles of NSCs and SBHs, and the mass scales at which transitions occur, are therefore still only partially constrained.

SHARP-VESPER enables these tests by resolving NSCs on parsec scales in nearby low-mass systems, allowing age and metallicity gradients, internal rotation, and gas kinematics to be measured simultaneously. These observables make it possible to assess whether NSCs are currently growing in situ, whether they retain gas over multiple star-formation cycles, and whether their presence correlates with enhanced or suppressed SBH accretion.

At intermediate galaxy masses, the transition between NSC- and SBH-dominated nuclei can be identified through spatially resolved stellar kinematics. Changes in rotation, velocity dispersion, and higher-order LOSVD moments, combined with MICADO constraints on nuclear structure, allow the relative dynamical contributions of NSCs and SBHs to be quantified directly.

In massive early-type galaxies, distinguishing among different nuclear configurations requires detailed measurements of the central velocity-dispersion profile, higher-order stellar kinematics, and stellar-population uniformity. SHARP-VESPER makes it possible to probe these signatures directly and distinguish among different classes of compact nuclear components.
The observational signatures associated with these transitions remain poorly constrained. A systematic SHARP-VESPER survey will provide the uniform, parsec-scale measurements required to test these scenarios empirically.

\subsection{Local nuclear systems as analogues of early black-hole growth}

Recent JWST observations have revealed a population of compact, actively accreting nuclei at redshifts $z>4$, often identified as “Little Red Dots” (LRDs), where the rest frame optical colours and compact morphologies suggest a significant AGN contribution even when the extended stellar host is extremely faint or difficult to isolate photometrically \citep{Labbe2025,Matthee2024,Kokorev2024,Kocevski2025,Akins2025}. In parallel, JWST spectroscopy has provided direct evidence for rapidly assembled or strongly overgrown black holes relative to their hosts in individual systems at very high redshift, motivating renewed attention to the physical conditions that allow efficient SBH growth in the early Universe \citep{Maiolino2024,Juodzbalis2024}. While selection effects and modelling degeneracies remain important, these results collectively indicate that efficient black hole growth can occur at early times in environments where the stellar component is either still assembling, extremely compact, or observationally subdominant.

Interpreting these high-redshift systems benefits from isolating, in nearby galaxy nuclei and at high physical resolution, the mechanisms that regulate the coupled evolution of SBHs and compact stellar structures. Relevant local targets include systems with prominent NSCs, weak or absent stellar nuclei, and nuclei dominated by the SBH potential.

In this context, SHARP-VESPER provides the observational means to identify and characterise the diversity of local nuclear structures, including dense NSCs, weak or absent compact stellar nuclei, and compact stellar overdensities potentially associated with black-hole recoil.

SHARP-VESPER can place these nuclear structures within a common observational framework, enabling direct comparisons between different modes of nuclear assembly and evolution.

This approach does not rely on a one to one mapping between low-redshift nuclei and individual JWST sources; instead it provides empirical constraints that inform the interpretation of the emerging JWST population of compact high-redshift AGN candidates \citep{Labbe2025,Kocevski2025,Akins2025}.

\section{Scientific return and observational feasibility}

The primary outcome of SHARP-VESPER will be a homogeneous census of SBHs and NSCs in nearby galaxy nuclei. Dynamical measurements of SBH masses, NSC masses and internal kinematics, together with stellar-population diagnostics, will enable the stellar and black-hole contributions to the central potential to be quantified within the same systems.

The combined SHARP--MICADO data set will provide a direct observational test of the processes responsible for the assembly, modification, and survival of compact nuclear components. Different pathways for building galaxy nuclei can be distinguished observationally, including recurrent in situ star formation, star-cluster infall, and SBH-driven dynamical evolution. The resulting constraints will inform theoretical models of nuclear growth, particularly with regard to gas retention, feedback, dynamical heating, and the emergence of SBH-dominated nuclei.

The same observations will also provide an empirical framework for interpreting compact high-redshift AGN candidates discovered with JWST. 
By establishing how SBHs and compact stellar structures are related in nearby galaxies, SHARP-VESPER will constrain the physical processes responsible for their assembly and long-term evolution.

\rev{Exposure-time estimates based on central surface brightnesses in the range $\mu_I = 15$--17 mag arcsec$^{-2}$ indicate that a signal-to-noise ratio of $S/N \simeq 30$ per spaxel (31$\times$31 mas$^2$), at $\lambda = 2.2\,\mu\mathrm{m}$ and a spectral resolving power of $R = 3000$, can be achieved with exposure times ranging from $180$ to $3000$ s. At larger radii, where the surface brightness is fainter, the required $S/N$ can be recovered through spatial binning. Even for a relatively faint surface brightness of $\mu_I = 20$ mag arcsec$^{-2}$, an $11\times11$ spaxel bin yields $S/N \simeq 30$ in an exposure time of $1.4$ h. This enables kinematic measurements in the regions sampled by the additional IFUs, while preserving the highest spatial resolution in the nucleus. Typical observations therefore require of order one hour of integration time per galaxy.}

\rev{The mass and distance regimes illustrated in Fig.~\ref{fig:euclid_double_panel} encompass a few hundred nearby galaxies with different masses, morphologies, and environments. Potential targets can be selected among nearby field and group galaxies \citep[e.g.][]{Cappellari2011}, as well as galaxies in nearby clusters such as Virgo and Fornax \citep[e.g.][]{Ferrarese2006_ACS,Jordan2007}.}

The resulting measurements will provide empirical constraints on the physical processes that shape galaxy nuclei.

%\ackheads
\section*{Acknowledgements}
All authors read and approved the submitted version of the paper.
EDB, AP, and EMC acknowledge the support of Padua University grants DOR 2025-2026.
EP acknowledges INAF Bando di Ricerca Fondamentale INAF 2022- MiniGrant RSN1.

\bibliographystyle{cas-model2-names}
\bibliography{sharp_nsc_bh_revised} 

\end{document}